# Revenge of the 'Neurds': Characterizing Creative Thought in terms of the Structure and Dynamics of Memory

Liane Gabora

University of British Columbia

Address for Correspondence:
L. Gabora <liane.gabora@ubc.ca>
Department of Psychology
University of British Columbia
Okanagan campus, 3333 University Way
Kelowna BC, V1V 1V7
CANADA

**Abstract**

There is cognitive, neurological, and computational support for the hypothesis that defocusing attention results in divergent or *associative thought*, conducive to insight and finding unusual connections, while focusing attention results in convergent or *analytic thought*, conducive to rule-based operations. Creativity appears to involve both. It is widely believed that it is possible to escape mental fixation by spontaneously and temporarily engaging in a more associative mode of thought. The resulting insight (if found) may be refined in a more analytic mode of thought. The questions addressed here are: (1) how does the architecture of memory support these two modes of thought, and (2) what is happening at the neural level when one shifts between them? Recent advances in neuroscience shed light on this. Activated cell assemblies are composed of multiple *neural cliques*, groups of neurons that respond differentially to general or context-specific aspects of a situation. I refer to neural cliques that *would not* be included in the assembly if one were in an analytic mode, but *would* be if one were in an associative mode, as *neurds*. It is posited that the shift to a more associative mode of thought is accomplished by recruiting neurds that respond to abstract or atypical microfeatures of the problem or task. Since memory is distributed and content-addressable, this fosters the forging of associations to potentially relevant items previously encoded in those neurons. Thus it is proposed that creative thought not by searching a space of predefined alternatives and blindly tweaking those that hold promise, but by evoking remotely associated items through the recruitment of neurds in a distributed, content-addressable memory.

What is happening in the brain when one engages in creative thought? The complex nature of creativity makes the goal of achieving a biological account of it difficult (Runco, 2004). However, progress has been made. Hoppe and Kyle (1991) studied Sperry's (1966) infamous commissurotomy patients (who underwent surgical bisection of the corpus callosum to inhibit epileptic seizures) and found they lacked the capacity for integrated thought and affect-laden interpretation of experience, which they speculated was related to these patients' "impoverished fantasy life". Their work suggested that interaction between the two hemispheres of the brain is important for creativity. Eysenck (1993) provides evidence that there is a genetic basis to creativity, and that creative individuals are statistically more likely than average to have relatives who are psychotic. Recent studies have found evidence that different kinds of creativity (deliberate versus spontaneous, and emotional versus cognitive) involve different neural circuits (Dietrich, 2004). Despite such accomplishments, a biological account of the creative process still eludes us.

This article synthesizes findings that together provide a tentative explanation for what is taking place at the neural level when one puts information together in a new and useful, creative manner. The paper begins by reviewing cognitive, biological, and computational evidence for the hypothesis that creativity involves the capacity to spontaneously shift back and forth between analytic and associative modes of thought according to the situation. Noting that memory is not just a storehouse of previous experiences, we then examine how mental representations are encoded as cell assemblies composed of neural cliques in a sparse, distributed, content-addressable memory, and how this cognitive architecture is navigated in a stream of thought. We will see why this kind of memory architecture enables us to implicitly know more than we have ever explicitly learned, and examine the role this likely plays in the capacity to generate creative ideas. Integrating research from cognitive psychology, neuroscience, and computational modeling, it is proposed that the shift to an associative mode of thought conducive to creative insight is accomplished by recruiting *neurds*: neural cliques that respond to abstract or atypical aspects of a problem, task, or situation. This fosters the forging of associations to potentially relevant items previously encoded to them. The paper concludes with a concrete example of what this theory predicts is happening at both the cognitive level and the neural level during a creative problem solving situation.

## Attributes of Creative Individuals

Theories of how the creative process works owe much to studies of characteristics associated with high creativity. Martindale (1999) identified a cluster of such attributes. One is *defocused attention:* the tendency to not focus exclusively on the relevant aspects of a situation, but notice also seemingly irrelevant aspects (Dewing & Battye, 1971; Dykes & McGhie, 1976; Mendelsohn, 1976). Finke et al. (1992) suggest that broadening the focus of attention might improve creativity and help overcome fixation—the mental state of individuals who are unable to move beyond a known problem solving approach to generate a new one (Jansson & Smith, 1991; Smith, 1995). A related attribute is high sensitivity (Martindale, 1977, 1999; Martindale & Armstrong, 1974), including sensitivity to subliminal impressions; stimuli that are perceived but of which one is not *conscious* of having perceived (Smith & Van de Meer, 1994). Other traits of creative individuals include risk taking, tolerance of ambiguity and delayed gratification, as well as strong leanings toward nonconformity and unconventionality (Crutchfield, 1962; Sternberg, 2006; Sulloway, 1996).

Creative individuals also tend to have *flat associative hierarchies* (Mednick, 1962). The steepness of one's associative hierarchy is measured by comparing the number of words generated in response to stimulus words on a word association test. Those who generate few words for each stimulus have a *steep* associative hierarchy, whereas those who generate many have a *flat* associative hierarchy. Thus, once such an individual has run out of the more usual associations (*e.g.,* CHAIR in response to TABLE), unusual ones (*e.g.,* ELBOW in response to TABLE) come to mind. The evidence that creativity is associated with both defocused attention and flat associative hierarchies suggests that creative individuals not only notice details others miss, but these details get etched in memory and are available later on. One's encoding of a situation includes aspects that are less central to the particular concept that best categorizes it, features that may in fact make it defy straightforward classification as strictly an instance of one thing or another.

However, a considerable body of research suggests that creativity involves not just the ability to defocus and free-associate, but also the ability to focus and concentrate (Eysenck, 1995; Feist, 1999; Fodor, 1995; Richards, Kinney, Lunde, Benet, & Merzel, 1988; Russ, 1993). Feist (1999) described how that creative people are, paradoxically, both labile and mutable yet controlled and stable. As Barron (1963) put it: "The creative genius may be at once naïve and knowledgeable, being at home equally to primitive symbolism and rigorous logic. He is both more primitive and more cultured, more destructive and more constructive, occasionally crazier yet adamantly saner than the average person" (p. 224). How do we make sense of this seemingly contradictory picture of the creative individual?

## Contextual Focus: Shifting Between Modes of Thought

This paradox can be reconciled. While some posit that creativity involves everyday thought processes such as remembering, planning, reasoning, and restructuring (Perkins, 1981; Weisberg, 2006a, 2000b), others argue that there are cognitive processes unique to creativity. These include the generation of unusual or 'remote' associations (Mednick, 1962) through processes such as bisociation (Koestler, 1964/1989), Janusian thinking (Rothenberg, 1976, 1979, 1982), integration (Sternberg, 1999), emotional resonance (Lubart & Getz, 1997) and divergent or associative thinking (Guilford, 1950, 1967; Khandwalla, 1993; Runco, 1991). Associative thought is said to be intuitive, unconstrained, and conducive to unearthing remote or subtle associations between items that share features or are *correlated* but not necessarily *causally* related. This may yield a promising idea or solution though perhaps in a vague, unpolished form. Extensive evidence suggests that associative thinking is correlated with controlled access to, and integration of, affect-laden material, or what Freud (1949) referred to as 'primary process' content (Russ, 1993, 2001). Associative thought is contrasted with a rule-based, convergent, or *analytic* mode of thought that is conducive to analyzing relationships of cause and effect between items already believed to be related. Analytic thought is believed to be related to what Freud termed 'secondary process' material.

There is an enduring notion that thought varies along a continuum between these two extremes, associative and analytic (Arieti, 1976; Ashby & Ell, 2002; Freud, 1949; Guilford, 1950; James, 1890/1950; Johnson-Laird, 1983; Kris, 1952; Neisser, 1963;

Piaget, 1926; Rips, 2001; Sloman, 1996; Werner, 1948; Wundt, 1896). This suggests a resolution to the seemingly contradictory characteristics of the creative individual. Several researchers have converged upon the view that creativity involves the ability to either shrink or expand the field of attention, and thereby match where one's mode of thought lies on the spectrum from associative to analytic according to the situation one is in (Finke et al., 1992; Gabora, 2000a, 2000b, 2003; Howard-Jones & Murray, 2003; Martindale, 1995). In associative thought one considers items in detail or considers multiple items at once, which facilitates detecting likenesses and integrating them. In analytic thought one considers items in a compact or 'atomic' form which facilitates mental operations on them. This capacity to shift between analytic and associative thought is sometimes referred to as *contextual focus*, since it is brought about by the situation or context (Gabora, 2003).[1] Related to contextual focus is the proposal that creativity occurs through interaction between productive and critical modes of thinking (Israeli, 1962, 1981), or ideation-evaluation cycles (Basadur, 1995). Although it has been assumed that contextual focus is subconscious, evidence that some degree of personal control may be involved comes from studies that showed that simply encouraging people to "be creative" increases scores on tests of divergent thinking (Harrington, 1975).

Note that to be constantly in a state of defocused attention, in which relevant dimensions of a situation do not stand out clearly from irrelevant ones, would be impractical. Since associative thought is of little value in many of the routine tasks of daily life, and since it can lead one's attention away from the 'here and now', it would be dangerous to have the capacity to engage in it unless one had the ability to stop it. It is only when one does not yet know what *are* the relevant dimensions—or when those assumed to be relevant turn out not to be—that defocused attention is of use. After the relevant dimensions have been found, it is efficient to focus on them exclusively. Associative thought would be adaptive only if the capacity for it were to have evolved side-by-side with the capacity to revert to a more analytic mode of thought when needed (i.e. if some pressing situation arises that demands a quick, logical response). Once the capacity to shift between analytic and associative modes of thought arose as needed, however, the capacity for creativity would be unprecedented. The explosion of creativity in the Middle/Upper Paleolithic has led to speculation that the capacity for contextual focus arose at this time (Gabora, 2003; Gabora & Aerts, 2009).

In sum, when the creator is stuck or fixated, and progress is not forthcoming, attention becomes more defocused, the creator enters a more associative mode of thought, such that peripherally related elements of the situation are more readily considered. This may continue until a potential solution is glimpsed, at which point attention becomes more focused, and thought becomes more convergent, as befits the fine-tuning and manifestation of the creative work. Noting that a creative idea is generally considered to possess two main qualities—appropriateness and originality—Howard-Jones and Murray (2003) suggested that divergent thought ensures originality while convergent thought ensures appropriateness.

**Biological Support for Contextual Focus**

There is indirect neuroscientific support for contextual focus, and in particular for the notion that creativity involves the ability to match where one's mode of thought lies on the spectrum from associative to analytic according to the situation one is in. Prior to finding the solution to an insight problem there is neural recruitment of the prefrontal and executive memory networks, as well as the so-called 'default network' associated with spontaneous mind wandering (Christoff, Gordon, Smallwood, Smith, & Schooler, 2009; Christoff, Gordon, & Smith, 2008; Kounios et al., 2006, 2008). This suggests that mind wandering has a utilitarian function, and provides neurological support for the notion of expanded receptivity through neural recruitment during associative thought.

There is evidence of an association between creativity and high variability in physiological measures of arousal such as heart rate (Bowers & Keeling, 1971; Jausovec & Bakracevic, 1995), spontaneous galvanic skin response (Martindale, 1977), cortical activity (Martindale & Armstrong, 1974) and EEG alpha amplitude (Hoppe & Kyle, 1991; Martindale, 1999; Martindale & Hasenfus, 1978). For example, although creative people tend to have higher *resting* arousal levels, low arousal, as measured by EEG (percentage of time spent in alpha states), is correlated with more creative problem solving (Martindale & Armstrong, 1974). Interestingly, low cortical arousal is observed only during the inspiration phase or divergent thinking component of the creative process. Divergent thought appears to be facilitated by lower levels of noradrenaline and dopamine—catecholamines directly linked to cognitive control, prefrontal functioning, and cortical arousal (for review see Heilman, Nadeau, & Beversdorf, 2003). Electroencephalography (EEG) experiments show that divergent thinking tasks produce decreased beta range synchrony and increased alpha range synchrony over the frontal cortex, providing further evidence for a loosened cognitive control and lower prefrontal cortical arousal during creative thought (Fink & Neubauer, 2006; Molle et al., 1996; Molle, Marshall, Wolf, Fehm, & Born, 1999; Razoumnikova, 2000, 2007). Collectively these findings suggest that during creative problem solving, creative individuals are particularly prone enter a state that is quite different from their normal resting state, a state that has both a physiological aspect (e.g. low arousal level) and a cognitive aspect (more divergent mode of thought).

**Computer Models Relevant to Contextual Focus**

The notion that thinking involves shifting between analytic and associative modes has been explored computationally. Starting with a model of associative memory it is possible, though not efficient, to simulate associative thought by injecting randomness. Martindale (1995) points out that something very much like a shift between divergent and convergent thought takes place in *simulated annealing* in a Hopfield network (Hopfield, 1982).[2] A Hopfield network is a kind of neural network that borrows the concept of energy minimization from physics. In the context of a Hopfield network, the term energy minimization refers to the degree of constraint satisfaction in a set of connected nodes, *i.e.,* the higher the degree of constraint satisfaction the lower the energy. The extent to which nodes activate one another depends on the weights of the links connecting them, which varies according to a probabilistic function. The term *temperature* is used to refer to the degree of randomness in the weights. At a low temperature, nodes tend to activate only their adjacent neighbors, and to a predictable degree, while at a high temperature they behave more erratically. It could be said that high temperature is like divergent thought whereas low temperature is like analytic thought.

A similar procedure was employed in Copycat, an analogy solving computer program (Mitchell, 1993; Mitchell & Hofstadter, 1989). Here temperature modulates the degree to which not just typical but atypical associations are made. Therefore the activation of nodes at a high temperature does not come from 'out of the blue'; it reflects genuine associative structure, but is not restricted to those associations that are strongest. This is closer to the behavior of creative individuals in Mednick's (1962) associative hierarchies studies who give ELBOW in response to TABLE, which has *some* relevance to TABLE, just not nearly as much as CHAIR. The point is that the escape from fixation or 'breaking out of a rut' is accomplished not by injecting randomness, as shown in Figure 1a, but by capitalizing on subtleties in the associative structure of the network, as shown in Figure 1b.

[INSERT FIGURE 1 ABOUT HERE]

In Copycat, the loosening of associations with increased temperature is not dependent on context. The halo of concepts activated by a given concept simply gets wider, in the same predictable way, no matter what the particular analogy problem is, or what other concepts are activated. In humans, however, a word or concept's implicitly activated associative structure *is* clearly linked to and dependent upon context (Barsalou, 1982; Nelson, Goodmon, & Ceo, 2007). In the context 'kitchen table' one might give ELBOW as an associate of TABLE, but not in the context 'pool table', and certainly not in the context 'multiplication table'. (As is customary, concepts are indicated in capital letters.) As another example, shown in Figure 1c, under the context 'Christmas', a typical exemplar of TREE is FIR and a typical property is 'needles', but not so under the context 'desert island'. The generation of associations in creative thought is biased by context to lead to associates that are relevant to the problem (albeit in potentially obscure ways), and liable to lead to a solution. How is this seemingly magical feat achieved?

A hint as to how this might be accomplished comes from stimulus classification tasks, which have shown that psychological space is stretched along context-relevant dimensions and shrunk along context-irrelevant dimensions (Nosofsky, 1987). In ALCOVE, a computer model of category learning, only when activation of each input unit is multiplied by an attentional gain factor does the output match the behavior of human participants (Kruschke, 1992; Nosofsky & Kruschke, 1992). Thus learning and creative problem solving involve shifts in attention that determine how to structure the region of conceptual space held in working memory given the situation one is in, and the process of ongoing restructuring highlights those properties of a concept that are particularly relevant in the given context. The context-dependent restructuring of the problem or task enables the generation of associates that are biased by context to be more likely than chance to be relevant to the problem (albeit in potentially obscure or unknown ways). This kind of context-sensitivity is consistent with evidence that context, both physical (McCoy & Evans, 2002) and social (Amabile, 1996; Csikszentmihalyi, 1996; Feldman, 1999; Feldman, Csikszentmihalyi, & Gardner, 1994; Howard-Jones & Murray, 2003; Perkins, 2000; Seifert, Meyer, Davidson, Patalano, & Yaniv, 1995; Sternberg, Kaufman & Pretz, 2002) plays an important role in creative thought. However the activation of associates in creative thought reflects not just the current context but one's history of previous contexts, and those of which are relevant to the current task may come together in a new way to produce a new idea. In the next section we dig a little deeper to see how this could work.

## The Architecture of Memory

We now examine evidence that an associative memory contains information that was never explicitly stored there but that is implicitly present nonetheless due to the ingenious way one's history of experiences is encoded. It is proposed that this information is accessed in associative thought, and made increasingly explicit in analytic thought, enabling one to go beyond what one knows without resorting to trial and error.

We take as a starting point some fairly well established characteristics of memory. Human memories are encoded in neurons that are sensitive to ranges (or values) of *microfeatures* (Churchland & Sejnowski, 1992; Churchland, Sejnowski, & Arbib, 1992; Smolensky, 1988). For example, one might respond to a particular shade of red, or the quality of being shrewd, or quite likely, something that does not exactly match an established term (Miikkulainen, 1997). Although each neuron responds maximally to a particular microfeature, it responds to a lesser extent to related microfeatures, an organization referred to as *coarse coding* (Hubel & Wiesel, 1965). Not only does a given neuron participate in the encoding of many memories, but each memory is encoded in many neurons. For example, neuron *A* may respond preferentially to lines of a certain angle (say 90 degrees), while its neighbor *B* responds preferentially to lines of a slightly different angle (say 91 degrees), and so forth. However, although *A* responds *maximally* to lines of 90 degrees, it responds somewhat to lines of 91 degrees. The upshot is that an item in memory is *distributed* across a cell assembly that contains many neurons, and likewise, each neuron participates in the storage of many items (Hebb, 1949; Hinton, McClelland, & Rumelhart, 1986). A given experience activates not just *one* neuron, nor *every* neuron to an equal degree, but activation is spread across members of an assembly. The same neurons get used and re-used in different capacities, a phenomenon referred to as *neural re-entrance* (Edelman, 1987). Items stored in overlapping regions are correlated, or share features. Memory is said to be *content addressable;* there is a systematic relationship between the state of an input and the place it gets encoded. As a result, episodes stored in memory can thereafter be evoked by stimuli that are similar or 'resonant' in some (perhaps context-specific) way (Hebb, 1949; Marr, 1969).

This kind of distributed, content-addressable memory architecture is schematically illustrated in Figure 2. Each circle represents a microfeature that is maximally responded to by a particular neuron. Circles that are close together respond to microfeatures that are similar or related. The large, diffuse region of whiteness indicates the region of memory activated by the current thought or experience. Note that even if a brain does not possess a neuron that would respond maximally to the particular microfeature because its representations are distributed across *many* neurons, the brain is still able to encode that experience.

[INSERT FIGURE 2 ABOUT HERE]

The fact that memory is distributed and content-addressable is critically important for creativity. If it were not distributed, there would be no overlap between items that share microfeatures, and thus no means of forging an association between them. If it were not content-addressable, associations would not be meaningful. Content addressability is why the entire memory does not have to be searched or randomly sampled; it ensures that one naturally retrieves and blends items that are relevant. Content addressability also facilitates the activation of one item by another that is related to it in a rarely noticed but useful or appealing way. Recall that if the regions in memory where two distributed representations are encoded overlap then they share one or more microfeatures. They may have been encoded at different times, under different circumstances, and the correlation between them never explicitly noticed. But the fact that their distributions overlap means that *some* context could come along for which this overlap would be relevant or useful, and cause one to evoke the other. Content addressability also means that there are as many routes to an association or reminding event as there are microfeatures by which they overlap; *i.e.,* there is plenty of room for typical as well as atypical connections to be made. Because the region of activated memory locations falls midway between the two extremes—not distributed and fully distributed—one can generate a stream of coherent yet potentially creative thought (Gabora, 2002). The more detail with which items have been encoded in memory, the greater their potential overlap with other items, and the more retrieval routes for creatively forging relationships between what is currently experienced and what has been experienced in the past.

Note how this differs from a typical computer memory. In a computer memory each possible input is stored in a unique address. Retrieval is thus a matter of looking at the address in the address register and fetching the item at the specified location. Since there is no *overlap* of representations, there is no means of creatively forging new associations based on newly perceived similarities. Even a simple connectionist memory is able to abstract a prototype, fill in missing features of a noisy or incomplete pattern, or create a new pattern on the fly that is more appropriate to the situation than anything it has ever been fed as input (Rumelhart & McClelland, 1986). The exceptions are by and large computer architectures that are designed to mimic, or are inspired by, human memory (Kanerva, 1988).

The upshot of all this is that there is no need to posit numerous variant solutions generated through chance processes. In a sparse, distributed, content-addressable memory, items that share features can, given the appropriate context, access one another even if their relationship has never been explicitly noted. Experimental support for the hypothesis that this is what actually happens in human cognition was obtained using the Accumulated Clues Task, or ACT (Bowers, Farvolden, & Mermigis, 1995). Participants were presented 15 clue words in succession, each of which was a weak associate to the solution word. After each clue, participants wrote down their guess as to what the solution word was. Independent of both where each clue was in the sequence of clues given, and the number of clues needed to solve the ACT, the semantic relatedness between guesses and the solution word increased linearly as a function of the number of clues. Thus participants were closer to the correct answer than chance prior to being able to actually *give* this correct answer. This supports the position that prior to insight, an answer is not waiting in a dormant, predefined state to be selected from amongst a set of others and tweaked or mutated in a trial and error manner to achieve its final form. Rather, the answer appears to emerge through the retrieval of associations in an initially vague or 'half-baked' form and over time become increasingly correct, appropriate, or well-defined. Actually, according to Edelman (2000), one does not retrieve a stored item from memory so much as *reconstruct* it. That is, an item in memory is never re-experienced in exactly the form it was first experienced, but colored, however subtly, by what has been experienced in the meantime, re-assembled spontaneously in a way that relates to the task at hand. Accessing and reflecting on this implicitly present information enables one to 'go beyond what one explicitly knows' to solve a problem (or simply express oneself) far more efficiently than trial and error.

Now we ask: how much overlap of microfeatures must there be for a creative association to be made? Another way of asking this is: how distributed must the memory be? At one extreme it could be not distributed at all, like a typical computer memory. If the mind stored each item in just one location as a computer does, then in order for an experience to evoke a reminding of a previous experience, it would have to be *identical* to that previous experience. And since the space of possible experiences is so vast that no two ever *are* exactly identical, this kind of organization would be fairly useless. But at the other extreme, in a *fully distributed* memory, where each item is stored in every location, the items interfere with one another. This phenomenon goes by many names: 'crosstalk', 'superposition catastrophe', 'false memories', 'spurious memories' or 'ghosts' (Feldman & Ballard, 1982; Hopfield, 1982; Hopfield, Feinstein, & Palmer, 1983).

The problem of crosstalk is solved by *constraining* the distribution region. One way to do this in neural networks is to use a radial basis function, or RBF (Hancock, Smith, & Phillips, 1991; Holden & Niranjan, 1997; Lu, Sundararajan, & Saratchandran, 1997; Willshaw & Dayan, 1990). Each input activates a hypersphere (sphere with more than three dimensions) of locations, such that activation is maximal at the center $k$ of the RBF and tapers off in all directions according to a (usually) Gaussian distribution of width $\sigma$. The result is that one part of the network can be modified without affecting the capacity of other parts to store other patterns. A *spiky activation function* means that $\sigma$ is small. Therefore only those locations closest to $k$ get activated, but they are activated a lot. A *flat activation function* means that $\sigma$ is large. Therefore locations relatively far from $k$ still get activated, but no location gets *very* activated.

The distinction between flat and spiky activation functions in neural networks is clearly reminiscent of Mednick's (1962) flat and steep associative hierarchies, and the notion that flat hierarchies are associated with associative thought while steep hierarchies are associated with analytic thought. Recall that contextual focus entails not just the capacity for both associative and analytic thought, but the capacity to adjust the mode of thought to match the demands of the problem at a given instant. There is reason to believe that we engage in contextual focus using a mechanism akin to varying the size of the RBF: spontaneous tuning of the spikiness of the activation function in response to the situation. With flat activation, items are evoked in detail, or multiple items are evoked at once, items with overlapping distributions of microfeatures. Thus flat activation is conducive to forging associations amongst potentially amongst items not usually thought to be related, or detecting relationships of correlation. With spiky activation, items are evoked in a compressed form, and few are evoked at once. Thus it is conducive to mental operations on those items, or deducing relationships of causation.

## Recruitment of Neurds in Associative Thought

Having examined the architecture of memory we have seen why one need not invoke randomness to explain how disparate elements are brought together in the creative process. Unusual ideas can come about through exploitation of uniquely forged associative structure, and this can be accentuated by associative thought. Let us now investigate more precisely how this might work.

It has been found that the cell assembly involved in the encoding of a particular experience is made up of multiple groups of collectively co-spiking neurons referred to as *neural cliques* (Lin et al., 2005; Lin, Osan, & Tsien, 2006). New techniques enabling their patterns of activation to be mathematically described, directly visualized, and dynamically deciphered, reveal that some cliques respond to situation-specific elements of an experience (*e.g.,* where it took place), while others respond to characteristics of varying degrees of abstractness or generality. These range from the type of experience (*e.g.,* being dropped) to characteristics common to many types of experience (*e.g.,* anything dangerous). Lin et al. depict this as a pyramid in which cliques that respond to the most context-specific elements are at the top of the pyramid, and those that respond to the most general elements are at the bottom.

What is a reasonable hypothesis for what is happening at the level of neural cliques during creative thought? Each successive thought activates recruitment of more or fewer neural cliques, depending on the nature of the problem, and how far along one is in solving it. Two well-established phenomena help ensure that this proceeds smoothly. First, if the same neurons are stimulated repeatedly they become refractory. For the duration of this refractory period they cannot fire, or their response is greatly attenuated. Second, they 'team play'; a response is produced by a cooperative group of neurons such that when one is refractory another is active. Since the situation-general neurons and the situation-specific neurons are not responding to the same aspects of the situation, they are not entering and leaving their refractory periods in synchrony, making it highly unlikely that one would think the same, identical thought over and over again (although over a longer time frame one may repeatedly cycle back to it).

Returning to Figure 2 we can get a schematic picture of how memory is activated by a particular thought. Recall that the degree to which any given region of memory is activated by the current thought or experience is indicated by the degree of whiteness. The white area thus represents the active cell assembly composed of one or more neural cliques, indicated by dashed gray circles. The further a neuron is from the center of the white region, the less activation it not only *receives* from the current instant of experience but in turn *contributes* to the next instant, and the more likely its contribution is cancelled out by that of other simultaneously active locations. Using neural network terminology, we say the broader the region affected by a given stimulus, the flatter the activation function, and the narrower the affected region, the spikier the activation function. Figure 2 portrays a situation in which the problem is the need to invent a comfortable, informal chair. The white region is narrow because it is activated in an analytic mode of thought. Thus only neurons that respond to very typical chair features such as 'has back' and 'has legs' are activated.

In a state of defocused attention more aspects of a situation get processed; the set of activated microfeatures is larger, and thus the set of potential associations one could make is larger. This situation is portrayed in Figure 3. The problem is still the need to invent a comfortable, informal chair, but here the activation function is flat. Recruitment of neural cliques that respond to abstract elements of the current thought (comfortable, informal) causes activation of an item in memory that share these abstract elements (beanbag). Its properties ('filled with stuffing' and 'conforms to shape') may *seem* irrelevant to the task at hand, but they turn out to play a key role in the invention of the beanbag chair.

[INSERT FIGURE 3 ABOUT HERE]

The neural cliques that do not fall within the activated region in Figure 2 but do fall within the activated region in Figure 3 are cliques that *would not* be included in a cell assembly if one were in an everyday relatively convergent mode of thought, but *would* be included if one were in an associative mode of thought. Let us refer to them as *neurds*. Neurds respond to microfeatures that are marginally relevant to the current thought. There is no particular portion of memory where neurds reside. The subset of neural cliques that count as neurds is defined by context, and shifts constantly. For each situation one might encounter, and for each new interpretation of that situation, a different group of neurds is involved.

The explanation proposed here for what happens in creative thinking follows naturally from the discovery of neural cliques that respond to varying degrees of specificity or generality, and the evidence for contextual focus (both outlined above), as well as the well-established phenomenon that activation of an abstract or general concept causes activation of its instances through spreading activation (Anderson, 1983; Collins & Loftus, 1975).[3] Given evidence that some neural cliques respond to specific aspects of a situation and others respond to more general or abstract aspects, we have a straightforward mechanism by which contextual focus could be executed. In associative thought, with more aspects of a situation taken into account, more neural cliques are activated, including those responding to specific elements, those responding to abstract elements, and those *they* activate through spreading activation. Activation flows from the specific instance to the abstract elements it instantiates, to other instances of those abstract elements. The concept of neurds thus provides a way of referring to those neural cliques that respond to features of these other instances that are not features of the original instance.

In the course of routine life, neurds are excluded from the activated cell assembly. Their time in the limelight comes when one has to break out of a rut. In associative thought, broad activation causes more neural cliques to be recruited, including neurds. This allows the next thought to stray far from the one that preceded it while retaining a thread of continuity. The associative network can be not just penetrated deeply, but traversed quickly, and there is greater potential for representations to 'bleed' into one another in ways they never have before. Thus the potential to unite previously disparate ideas or concepts is high.

## Analysis of a Creative Act

We have examined the relationship between contextual focus and the structure of human memory. This synthesis will now be applied to the analysis of a creative act. In keeping with the view that everyone is creative (Beghetto & Kaufman, 2007; Gardner, 1993; Runco, 2004), the creative act that we analyze is not an earthshaking achievement but a simple event in the life of an everyday person.

The situation that motivates the creative act is the following. Jane, a ski instructor, wants to build a fence so that she will feel safer and can let her dog run around in the yard. However, she cannot afford a new fence.

Jane paces the yard trying to solve the situation through a straightforward deductive process. Neural cliques that have encoded memories of particular fences, and that respond to the concept 'fence', are activated. Her inability to solve the problem rationally eventually leads to a spontaneous and subconscious defocusing of attention. She enters an associative mode of thought, and her activation function becomes flat, such that the associative structure of her memory is more widely probed. Characteristics of fence posts, such as 'tall', 'skinny', and 'sturdy', are still strongly activated, but now they become less tied to fences. As neurds get recruited, her memory begins to respond to not just the context-specific aspects but also the abstract, conceptual aspects of her situation *i.e*. moving further down Lin et al.'s (2006) feature-encoding pyramid. She starts thinking not just about different kinds of fences but different ways of safeguarding her property, and things that are fence-*like*.

Because memory is content-addressable (which as we saw earlier means that there is a systematic relationship between the content of an item and the locations in memory it activates), representations other than the concept 'fence' or memories of fences that have been encoded to these locations in the past come to mind. The neural cliques that respond to 'tall', 'skinny' and 'sturdy', now activated in the context of needing to build a fence, had previously encoded numerous memories of skis. These neural cliques spread activation to cliques that respond to more specific items that are 'tall', 'skinny' and 'sturdy', such as skis, resulting in a combining of the concepts 'skis' and 'fence' to give a new concept: a fence made of skis. SKI FENCE has some properties unique to skis (*e.g.* bindings and curved tips) and some properties unique to fences (*e.g.* surrounds and protects property). This new association between SKIS and FENCE is not literally a connection but a distributed set of microfeatures that have never been activated together before as an ensemble.

Jane goes to a shed crammed full of old skis. Having hit upon this idea of using skis to build a fence, she must determine if it would really work in practice. Although in the short run a flat activation function is conducive to creativity, maintaining it would be impractical since the relationship between one thought and the next may be remote; thus a stream of thought lacks continuity. Access to obscure associations would at this point be a distraction. Thus, now that the framework of her idea has been painted in the broad strokes, she enters a more analytic mode by 'decruiting' the neurds, thereby narrowing the region of memory that gets activated. Thought becomes more logical in character because the activation function becomes spikier, thereby affording finer control; fewer locations release their contents to participate in the formation of the next thought. By focusing attention on the promising aspects of the idea (such as that skis are long and skinny and available) and ignoring irrelevant aspects (such as that skis have bindings) Jane figures out things like what to use as crossbeams and how to drive the skis into the ground. She thus settles on a workable solution to her problem.

In fact the situation is slightly more complex, because some aspects of adapting the idea of building a fence to using skis instead of fence posts probably require or lend themselves to a slightly more associative mode of thought. For example, perhaps white rocks at the base of the skis would not only help stabilize the skis but be suggestive of snow. By shifting back and forth along the spectrum from associative to analytic, the fruits of associative thought become ingredients for analytic thought, and *vice versa*. Notice how it was the nature of the problem or task constraints – the unattainable 'free fence' – that guided the entire process.

## Summary and Conclusions

This paper has attempted to synthesize experimental and theoretical work on creativity and memory into a reasonable account of how the brain generates novel ideas. A long-held view is that there are two modes of thought: (1) divergent or associative thought, which is conducive to unearthing similarity or relationships of correlation between items not previously thought to be related, and (2) convergent or analytic thought, which is conducive to hammering out causal relationships between items already thought to be related. It has been suggested that analytic thought requires a state of focused attention, and associative thought requires a state of defocused attention. Creativity involves not just the capacity for both, but the capacity to spontaneously shift between them according to the situation, referred to as contextual focus. Integrating evidence from cognitive science and neuroscience, it was suggested that contextual focus is explicable at the level of cell assemblies. A cell assembly is composed of neural cliques, some of which respond to situation-specific aspects of an experience, and others of which respond to general or abstract aspects (Lin et al., 2005, 2006). The contents of the activated cell assembly merge in the generation of the next instant of thought. It is proposed that the shift to a more associative mode of thought is accomplished by spontaneously flattening the activation function through the recruitment of neurds. Those neural cliques that respond to atypical features of the situation, and thus that are activated in associative but not analytic thought, are referred to as neurds.

Leakage of information from outside the problem domain occurs in a manner that is *a priori* unpredictable because it involves unearthing associations that exist due to overlap of items in memory that may not have been previously noticed. Because memory is distributed, coarse-coded, and content-addressable, items encoded previously to neurds are superficially different from the present situation yet share aspects of its deep structure. Therefore, the recruitment of neurds may foster associations that are seemingly irrelevant yet potentially vital to the creative idea. By responding to abstract or atypical features of the situation, neurds effectively draws new concepts into the conceptualization of the problem. Once an insight has been found, one finds a way for it to be realized by focusing attention, increasing the spikiness of the activation function, and dropping neurds from the activated cell assembly.

It is interesting to consider the long-term consequences of the proclivity to shift readily into a defocused state of attention. More aspects of attended stimuli participate in the process of encoding an instant of experience to memory and evoking 'ingredients' for the next instant of experience. The more *they* can in turn evoke, and so on. Thus not only is thought more associative, but streams of thought last longer. If something does manage to attract attention, it tends to be more thoroughly processed before settling into a particular interpretation of it. New information, including socially transmitted information, is less able to compete with this processing of previously acquired information that is already set in motion. Moreover, the answering of one question often generates new questions. The end result is that one's internal web of understandings is forged in a unique way, which is in turn reflected in its creative output.

Lin et al. (2005) claim "Conversion of activation patterns of these coding unit assemblies into a set of real-time digital codes permits concise and universal representation and categorization of discrete behavioral episodes across different individual brains" (p. 6125). Unfortunately, the large-scale recording technique they used with mice would not be suitable for human subjects. If it were, by comparing recordings of particularly creative and less creative subjects on creatively demanding tasks, the existence and behavior of neurds could in principle be scientifically verified and investigated. Neurds are generally withheld from participating in a stream of thought, but when they do come out they come out with a vengeance, and their contribution is sometimes revolutionary.

## Acknowledgments

I would like to thank Kalina Christoff, Jim Davies, Stefan Leijnen, Bruce Mathieson, Apara Ranjan, and Mark Runco for discussion and helpful comments on the manuscript. This research was funded by grants from the GOA Program of the Free University of Brussels, and the Social Sciences and Humanities Research Council of Canada (SSHRC).

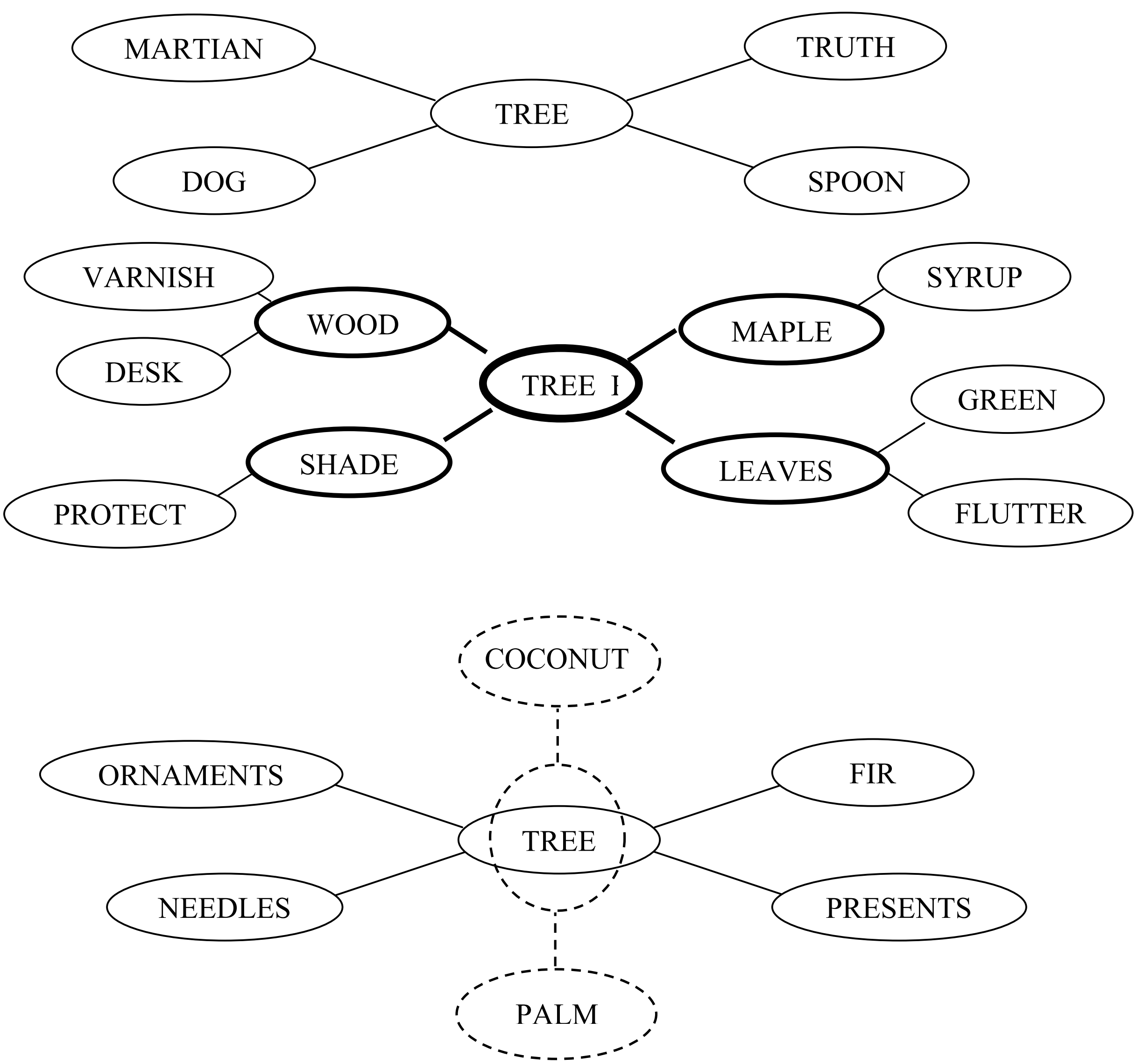

*Figure 1*. Three ways in which one concept could evoke recall of associated concepts. (a) Top: randomly generated associates. (b) Middle: generation of associates based shared properties or conceptual overlap. Diminished activation with distance from target (TREE) is indicated by narrower lines. (c) Bottom: context-sensitive generation of associates. Associates of TREE generated in the context 'desert island' (dashed lines) differ from those generated in the context 'Christmas' (un-dashed lines).

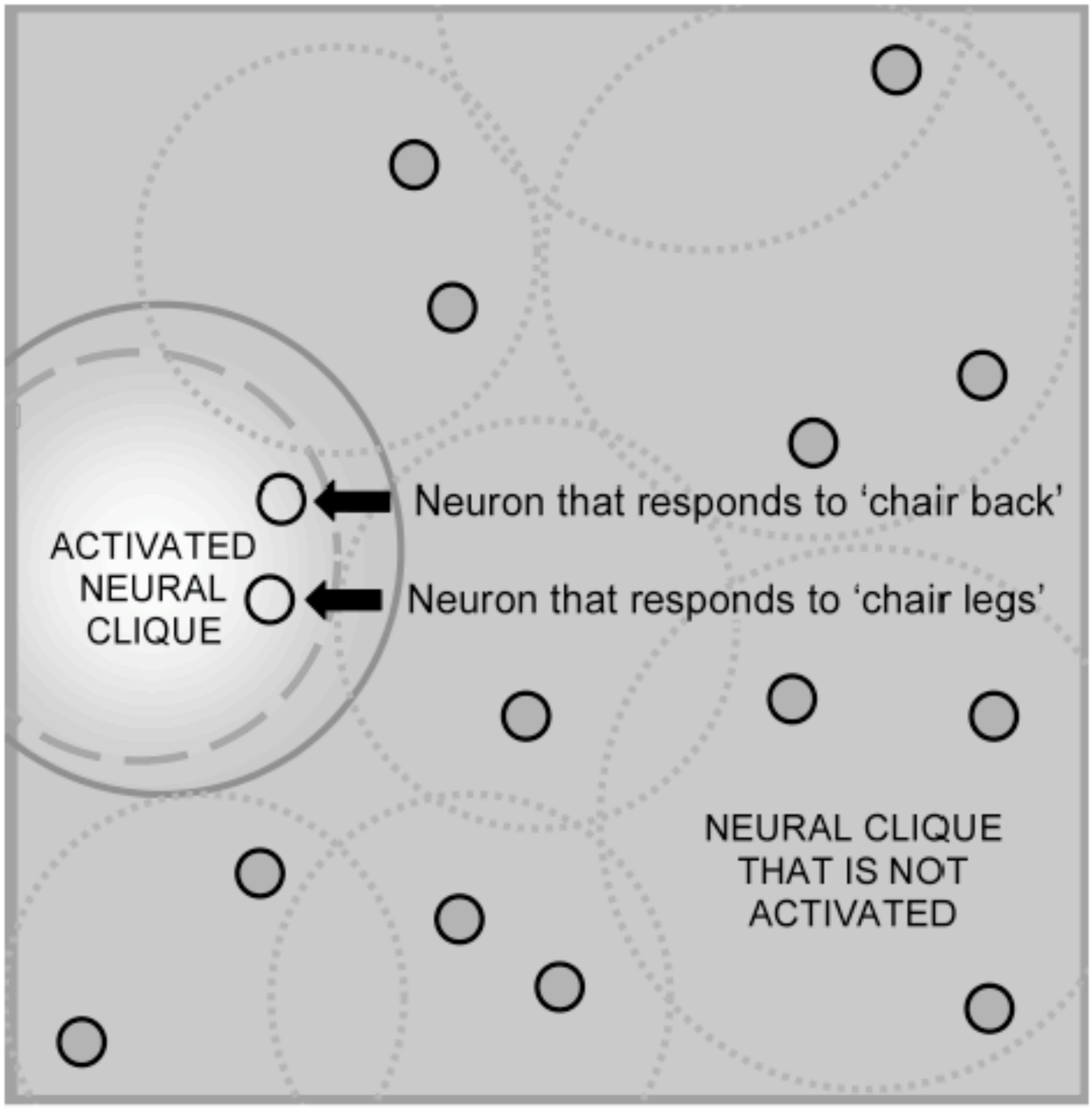

*Figure 2*. Schematized drawing of a portion of a distributed associative memory activated by the problem of inventing a comfortable chair in an analytic mode of thought. Each small black-ringed circle represents a feature that a particular neuron responds to. The white region indicates the portion of memory activated by problem. The activated cell assembly, indicated by the large grey circle, consists of only one neural clique, indicated by the dashed circle. It is composed of neurons that respond to typical features of chairs such as 'chair back' and 'chair legs'. Non-activated neural cliques are indicated by dotted gray circles.

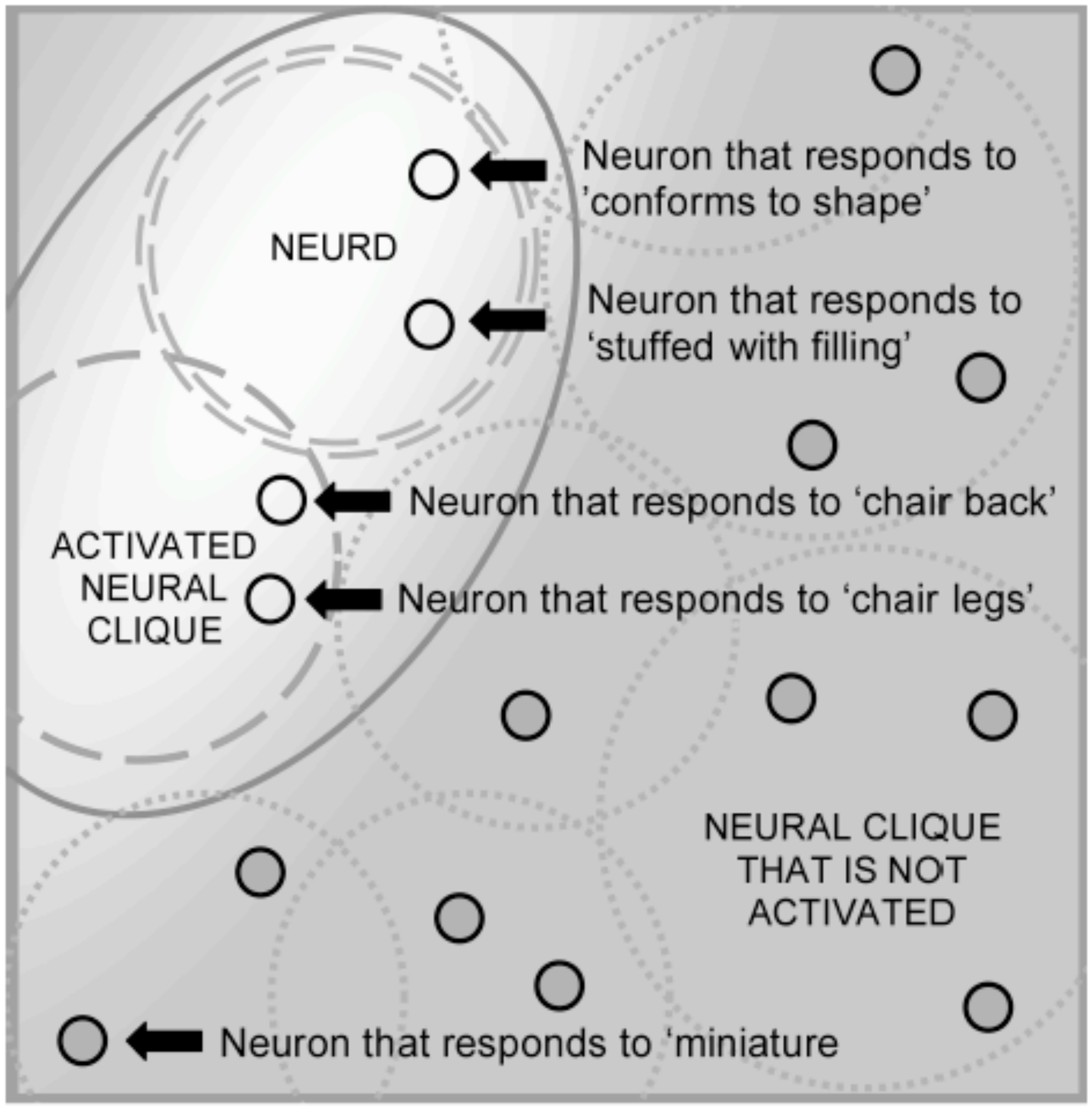

*Figure 3*. As indicated by the size and diffuseness of the white region, in an associative mode of thought, the portion of memory activated by the problem of inventing a comfortable chair is larger than it was in an analytic mode (Figure 2). The activated cell assembly, indicated by the large oval, now contains more than one neural clique. The initially activated neural clique is indicated by the dashed circle, and the neurd is indicated by the double circle of dashes. The neurd is composed of neurons that respond to atypical features of chairs such as 'conforms to shape' and 'filled with stuffing'. These features may nevertheless be relevant to the task of inventing a comfortable, informal chair, such as a beanbag chair. Note that under a different context, such as the task of making a chair for a doll, the neurd might have been a different neural clique, containing the neuron that responds to 'miniature'.

## Notes

[1] Note that divergent thought would be of adaptive value only if the capacity for it were to have evolved side-by-side with the capacity to revert to a more analytic mode of thought if needed (*i.e.*, if some pressing situation were to arise that demanded logical analysis and a quick response). Once the capacity to shift between analytic and associative modes of thought arose as needed, however, the capacity for creativity would be unprecedented. Thus it has been suggested that the explosion of creativity in the Middle/Upper Paleolithic was due to onset of the capacity for contextual focus at this time (Gabora, 2003, 2007).

[2] The term comes from the physical process of annealing in which one changes the properties of a metal by slowly lowering its temperature.

[3] Thus for example, based on a set of free association norms data collected from 6,000 participants using over 5,000 words, the probability that, given the word PLANET, the first word that comes to mind is EARTH is .61, and the probability that it is MARS is .10 (Nelson, McEvoy, & Schreiber, 2004). Note that there is some empirical support for an alternative to spreading activation as an explanation for this kind of association data, referred to as 'spooky activation at a distance' (Nelson, McEvoy & Pointer, 2003).